\documentclass[10pt]{article}
\usepackage{graphicx} %
\usepackage[colorlinks=true,urlcolor=urlblue,linkcolor=urlblue,citecolor=urlblue,breaklinks=true,bookmarks=false]{hyperref}
\usepackage[a4paper, total={5.75in, 9in}]{geometry}
\usepackage{amsmath}
\usepackage[sortcites, backend=bibtex]{biblatex}
\usepackage[all]{hypcap} %
\usepackage{matlab-prettifier}
\usepackage{titling}
\usepackage{xcolor}
\usepackage{authblk}

\newcommand{\clink}[2]{\href{#2}{\underline{\texttt{#1}}}}

\newcommand{\name}[1]{\textsc{VAE Explainer}}

\title{\name{}: Supplement Learning Variational Autoencoders with Interactive Visualization}
\author{Donald Bertucci and Alex Endert}
\date{\small Georgia Institute of Technology}

\begin{document}
\definecolor{urlblue}{RGB}{46,46,177}
\definecolor{fuscia}{RGB}{60,62,236}
\definecolor{limegreen}{RGB}{30,174,70}
\definecolor{horange}{HTML}{FFA500}
\definecolor{hseagreen}{HTML}{2E8B57}
\definecolor{hlightseagreen}{HTML}{20B2AA}
\definecolor{hsalmon}{HTML}{FA8072}
\definecolor{hlightblue}{HTML}{6fc7ec}

\definecolor{hpink}{HTML}{e44084}
\definecolor{hpurple}{HTML}{6821b0}
\definecolor{hgreen}{HTML}{90dc93}

\newcommand{\csigma}{\textcolor{hlightseagreen}{\sigma}}
\newcommand{\cmu}{\textcolor{horange}{\mu}}
\newcommand{\clogvar}{\textcolor{hseagreen}{\log(\sigma^2)}}
\newcommand{\ceps}{\textcolor{hsalmon}{\epsilon}}
\newcommand{\cz}{\textcolor{hlightblue}{z}}

\numberwithin{equation}{section}

\maketitle

\vspace{-15pt}
\noindent \textbf{Abstract:} 
Variational Autoencoders are widespread in Machine Learning, but are typically explained with dense math notation or static code examples. This paper presents \name{}, an interactive Variational Autoencoder running in the browser to supplement existing static documentation (e.g., \clink{Keras Code Examples}{https://keras.io/examples/generative/vae/}). \name{} adds interactions to the VAE summary with interactive model inputs, latent space, and output. \name{} connects the high-level understanding with the implementation: annotated code and a live computational graph. The \name{} interactive visualization is live at \clink{https://xnought.github.io/vae-explainer}{https://xnought.github.io/vae-explainer} and the code is open source at\\ \clink{https://github.com/xnought/vae-explainer}{https://github.com/xnought/vae-explainer}.

\section{Introduction}
Variational Autoencoders (VAE) \cite{kingma2013auto} compress data effectively and produce a latent space that can be nicely interpolated through. However, VAEs are conceptually more difficult than regular Autoencoders (i.e., Reparameterization) and are described with dense mathematical notation \cite{kingma2013auto}. Furthermore, documentation or notebooks on VAEs include code, but no live interactive exploration to show off key pieces of the VAE \cite{9051780, kerasvae, kangvaetut, xander, ibm, demistwei}.

\name{} doesn't aim to replace existing examples, but to supplement them with interactive visualization. \name{} specifically builds off of the demonstrated educational effectiveness of interactive explainers like CNN Explainer \cite{wang2020cnn}, Diffusion Explainer \cite{lee2023diffusion}, and Transformer Explainer \cite{cho2024transformer} but to explain VAEs.

\begin{figure}[h]
    \centering
    \includegraphics[width=1.0\linewidth]{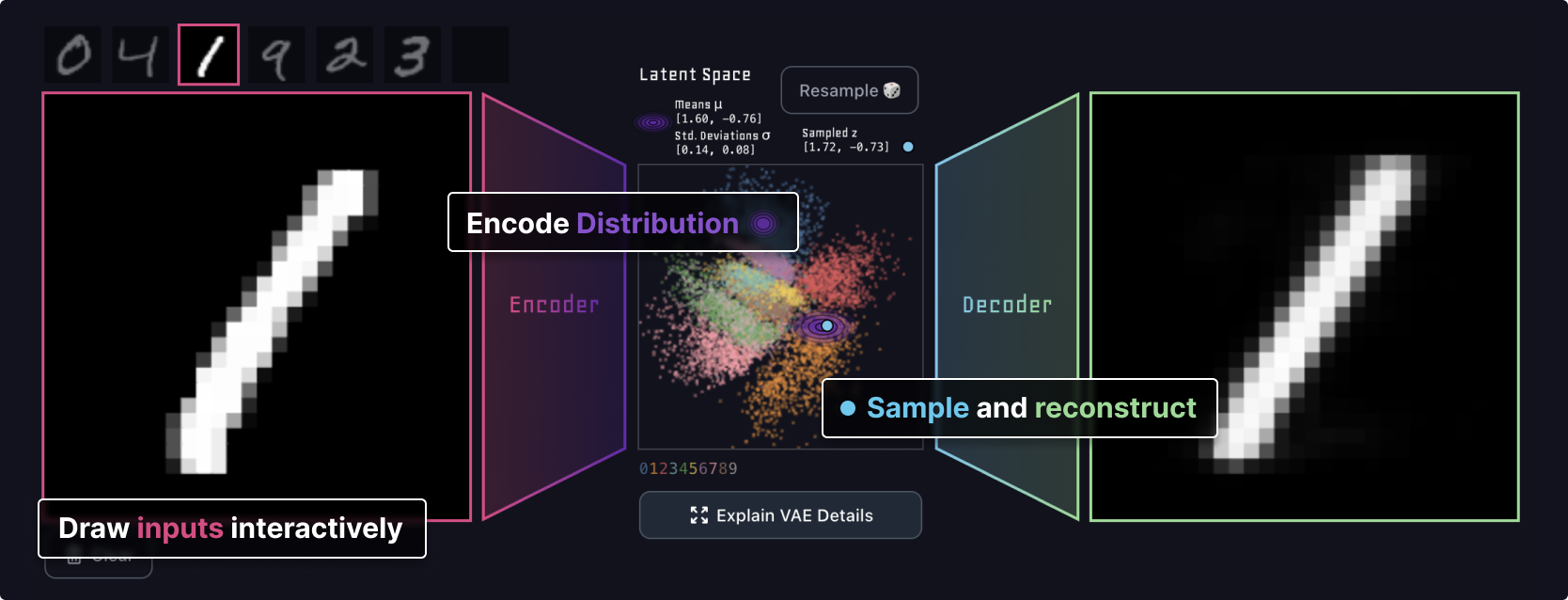}
    \caption{Users can draw a digit as \textcolor{hpink}{input} and the VAE runs in real-time. \name{} displays the \textcolor{hpurple}{encoded distribution} on top of the latent space. Then, we sample a \textcolor{hlightblue}{point} from the \textcolor{hpurple}{distribution} and decode into the \textcolor{hgreen}{reconstruction}.}
    \label{fig:teaser}
\end{figure}

\begin{figure}[t]
    \centering
    \includegraphics[width=1.0\linewidth]{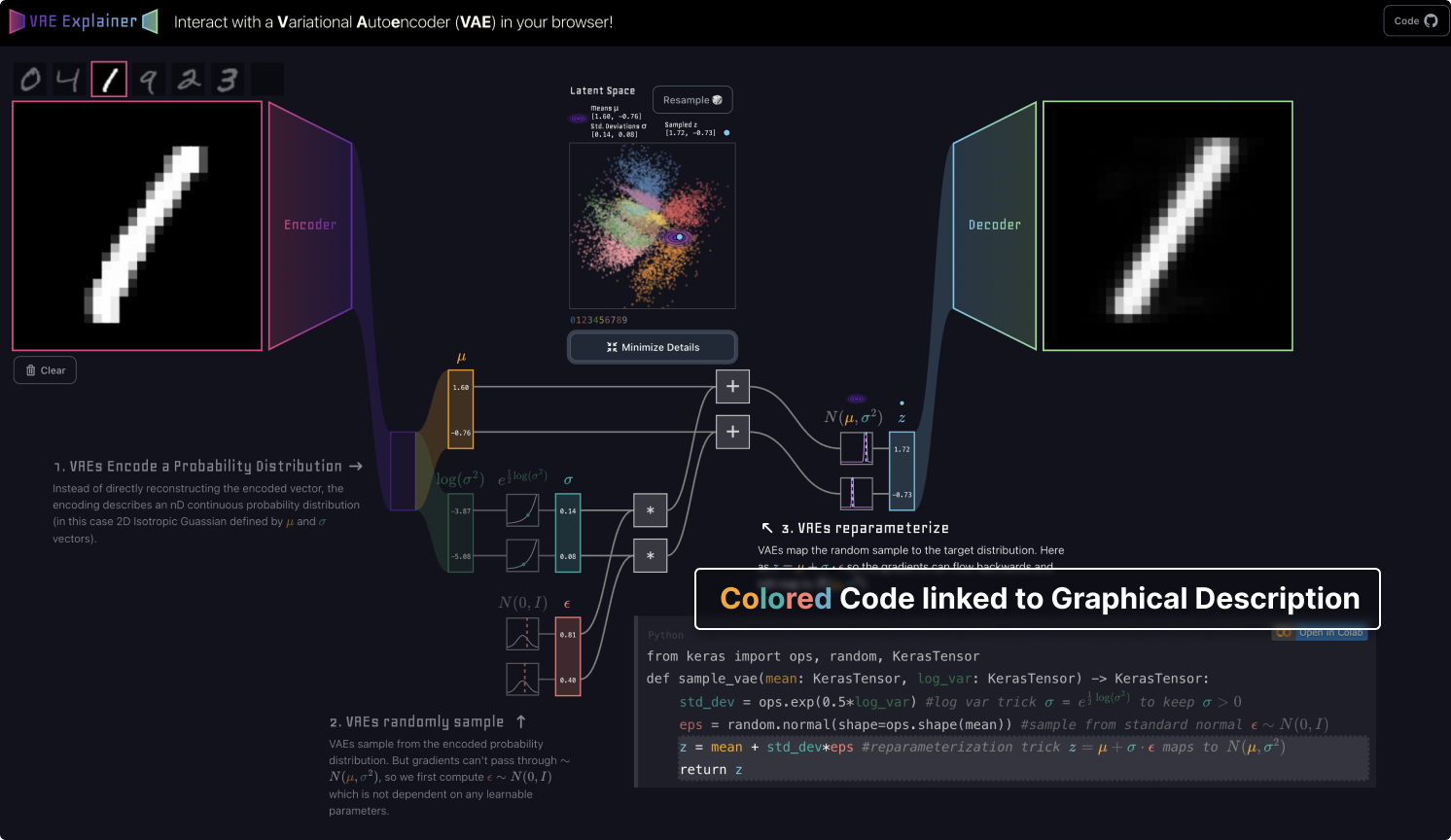}
    \caption{Users can click ``Explain VAE Details" to show annotated code connected to a computational graph. Hovering over lines in the code will highlight portions of the graph.}
    \label{fig:teaser2}
\end{figure}

 With \name{}, we don't display low-level details first. We hide the math notation and provide an interactive high-level overview (see \autoref{fig:teaser}). For example, a user can hand-draw the input and view how the encoded distribution and reconstruction changes. When a user is ready, they can display low-level implementation details such as the Log-Var Trick \cite{logvar} and Reparameterization Trick \cite{kingma2013auto} (see \autoref{fig:teaser2}). For simplicity and familiarity, we use the MNIST Digit dataset \cite{yann1998mnist} to align with existing documentation on VAEs \cite{kerasvae, ibm}.

\pagebreak{}
To be very specific, this paper contributes the following:
\begin{itemize}
    \item A high-level summary view of a VAE with interactive inputs and latent space (\autoref{subsec:overview}).
    \item A low-level graph view that describes implementation details (i.e., Log-Var \cite{logvar} and Reparameterization \cite{kingma2013auto} Tricks) with code and an annotated computational graph (\autoref{subsec:graph}).
    \item Open source and browser implementation to make \name{} accessible to anyone with a browser (\autoref{sec:impl}).
\end{itemize}

\section{System}
This section describes the entire \name{} tool in two subsections: the \textbf{High-Level Summary View} (\autoref{subsec:overview}) and the \textbf{Low-Level Graph View} (\autoref{subsec:graph}).

\subsection{High-Level Summary View}
\label{subsec:overview}
To explain the main ideas from static documentation on VAEs, \name{} distills the main point of a VAE as encoding a probability distribution of the input data, which we then sample and reconstruct (see \autoref{fig:teaser}).

\begin{figure}[t]
    \centering
    \includegraphics[width=1.0\linewidth]{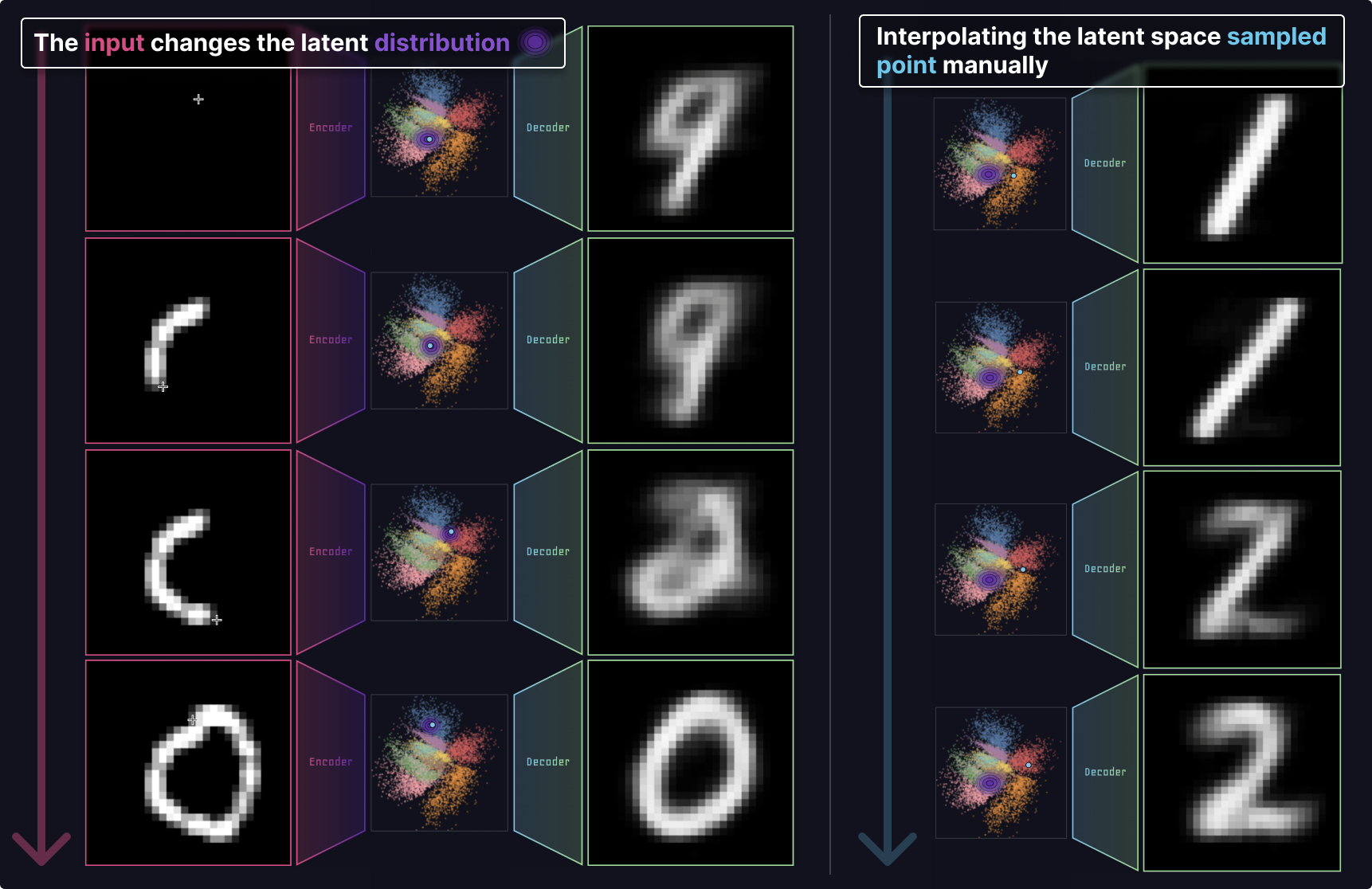}
    \caption{\textbf{Left side}: as we draw the digit ``0" in the \textcolor{hpink}{input}, the encoded distribution changes location and size to represent the distribution of possible ``0"s. \textbf{Right side}: as we hover the latent space and change the \textcolor{hlightblue}{sampled point}, we interpolate the reconstruction.}
    \label{fig:draw}
\end{figure}

The encoder takes a hand-written digit input and encodes the data as a two-dimensional isotropic normal distribution. We chose 2D so the latent space could be easily visualized by humans. The distribution itself is displayed directly on the latent space in gradually increasing and diffuse \textcolor{hpurple}{purple} circles (see middle of \autoref{fig:teaser}). Since the distribution has no covariance, it'll always be stretched in the vertical or horizontal direction. When you change the input data, you'll see that the distribution changes location and shape to other places in the latent space. For example on the left side in \autoref{fig:draw}, as we draw the digit ``0" as the input, the latent space gradually interpolates through ``9" and``2" regions before finding itself in the ``0" region. 

The latent space itself has many colored points in the background. These points are training data with labels from the MNIST dataset \cite{yann1998mnist}. When a user hovers over the latent space, they can change the sampled \textcolor{hlightblue}{blue} point to anywhere in space and see the reconstructed output. For example, on the right side in \autoref{fig:draw}, by hovering and moving the \textcolor{hlightblue}{blue} point from the ``1" region to the ``2" region in the latent space, we can see the interpolated reconstruction.   

\subsection{Low-Level Graph View}
\label{subsec:graph}
Once the user has a grasp of the overview, they can view the computations involved with the VAE by revealing the VAE computational graph as shown in \autoref{fig:teaser2}. This section connects the static documentation to the interactive pieces.

\begin{figure}[h]
    \centering
    \includegraphics[width=1.0\linewidth]{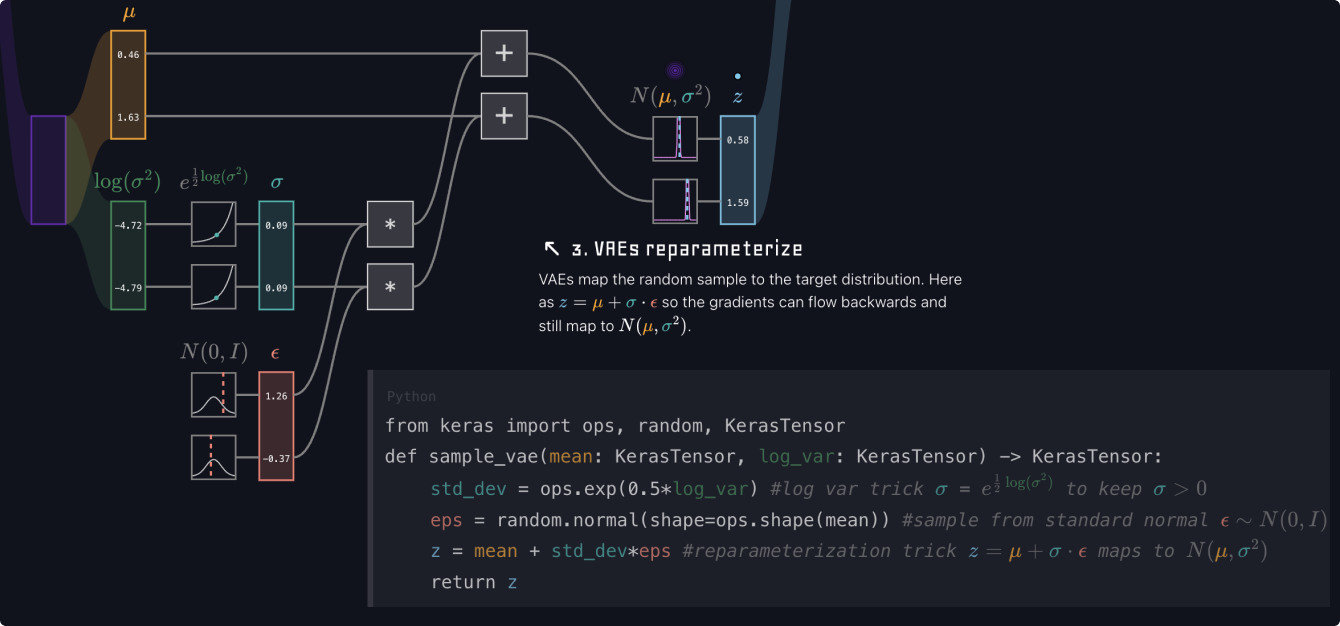}
    \caption{VAE sampling Keras code \cite{chollet2015keras} accompanied by its computational graph. Extra labels have been removed for figure presentation.}
    \label{fig:graph}
\end{figure}

First, the Keras \cite{kerasvae, chollet2015keras} Python code is displayed and colored so the notation is easier to understand \cite{color}. The code can be visualized as a computational graph as shown in \autoref{fig:graph}. We show the \textcolor{horange}{mean} vector $\cmu$ and the \textcolor{hseagreen}{log of the variances} vector $\clogvar$ with the real numbers on the graph. The encoder doesn't directly output the \textcolor{hlightseagreen}{standard deviation} $\csigma$ since the \textcolor{hlightseagreen}{standard deviation} must be greater than 0. Here we show the Log-Var Trick \cite{logvar} where we recover the $\csigma$ by applying 
\begin{align*}
    \csigma &= e^{\frac{1}{2}\clogvar}\\
    &= e^{\frac{1}{2}2\log(\csigma)}\\
    &= \csigma
\end{align*}
which forces the \textcolor{hlightseagreen}{standard deviation} $\csigma$ to be positive \cite{logvar}. The Log-Var trick is represented on the graph as mapping the encoding ($\clogvar$) through the exponential function node to produce the output ($\csigma$) vector (see \autoref{fig:graph}).

The parameters $\cmu$ and $\csigma$ specify the \textcolor{hlightblue}{normal distribution} $\cz \sim N(\cmu, \csigma^2)$ we sample from. The Reparameterization Trick \cite{kingma2013auto} samples $N(\cmu, \csigma^2)$ by sampling a \textcolor{hsalmon}{standard normal distribution} labeled as $\ceps \sim N(0, I)$ and mapped to $N(\cmu, \csigma^2)$ by
\begin{align*}
    \cz = \cmu + \csigma \cdot \ceps.
\end{align*}
The computational graph highlights the Reparameterization Trick \cite{kingma2013auto} by separating the $\ceps$ from the main pathway. A user can see that no parameters depend on $\ceps$ and that gradients can pass back to the encoder easily. In \autoref{fig:graph}, both probability distributions are shown as curves with vertical dotted lines to show values that are sampled. In \autoref{fig:teaser}, on the latent space the $\cz$ is the \textcolor{hlightblue}{blue} point and the $N(\cmu, \csigma^2)$ is the \textcolor{hpurple}{purple} distribution.

To make it completely obvious which code corresponds to what part of the graph, there is a two-way interaction. When a user hovers over a line of code, the graph highlights the corresponding computation and vice versa (see \autoref{fig:teaser2}).  

\section{Implementation}
\label{sec:impl}
To make \name{}, we trained an existing implementation of a VAE directly copied from the Keras Variational Autoencoder Example \cite{kerasvae} with some modifications for presentation. The training can be found in a \clink{Colab Notebook}{https://colab.research.google.com/github/xnought/vae-explainer/blob/main/notebooks/vae\_training\_keras\_cnn.ipynb}.

Just to summarize from \cite{kerasvae}, the model consists of a Convolutional Neural Network \cite{lecun2015deep} as the encoder and the opposite as the decoder (Convolution Transposes). The model was trained with the Adam optimizer \cite{kingma2014adam} over 30 epochs of the 60,000 MNIST Digits train set \cite{yann1998mnist}.

After training the model, we converted the Keras model to a TensorFlow graph and exported the graph to a TensorFlowJS format so it could be run in the browser \cite{tfjs, chollet2015keras, tensorflow2015-whitepaper}. We specifically exported the encoder and decoder as separate models so that the middle computation could be computed and visualized in the browser easily. Additionally, we computed the encodings for the first 10,000 MNIST Digit train set \cite{yann1998mnist} images to better map out the latent space in the browser.

We used JavaScript, TensorflowJS \cite{tfjs}, and Svelte \cite{svelte} for the interactive frontend. The visualizations are primarily SVG and Canvas elements. The frontend code can be found at the open source repository \clink{https://github.com/xnought/vae-explainer}{https://github.com/xnought/vae-explainer} and the live site can be found at \clink{https://xnought.github.io/vae-explainer}{https://xnought.github.io/vae-explainer}.

\section{Conclusion}
\name{} adds live interaction to static explanation. First a user can summarize what a VAE does, then they can view the real code and computational graph for how a VAE works.

To improve this work, more explanation on the VAE loss function would further help someone understand how the encoded normal distributions are regularized to standard normal. Additionally, extending to Vector Quantized Variational Autoencoders (VQ-VAE) would cover the latest and greatest for Autoencoders.

\subsection*{Acknowledgments}
We thank Adam Coscia for valuable feedback on early versions of the interactive website. Thank you Adam!

\printbibliography

\end{document}